\documentclass[twocolumn]{aastex63}
\usepackage{comment}
\usepackage{amsmath}
\usepackage{graphics}
\usepackage{color}
\usepackage{threeparttable}
\usepackage{rotating}
\usepackage{dcolumn}
\usepackage{multirow}
\newcolumntype{d}[1]{D{.}{.}{#1}}

\received{}
\revised{}
\accepted{\today}
\graphicspath{{./}{figures/}}

\begin{document}
\title{Dense Core Collisions in Molecular Clouds: Formation of Streamers and Binary Stars}

\author{Yuta Yano}
\affil{Department of Astronomy, The University of Tokyo, Hongo, Tokyo 113-0033, Japan}
\affil{National Astronomical Observatory of Japan, 2-21-1 Osawa, Mitaka, Tokyo 181-8588, Japan}

\author{Fumitaka Nakamura}
\affil{Department of Astronomy, The University of Tokyo, Hongo, Tokyo 113-0033, Japan}
\affil{National Astronomical Observatory of Japan, 2-21-1 Osawa, Mitaka, Tokyo 181-8588, Japan}
\affil{The Graduate University for Advanced Studies
(SOKENDAI), 2-21-1 Osawa, Mitaka, Tokyo 181-0015, Japan}

\author{Shinichi.W.Kinoshita}
\affil{Department of Astronomy, The University of Tokyo, Hongo, Tokyo 113-0033, Japan}
\affil{National Astronomical Observatory of Japan, 2-21-1 Osawa, Mitaka, Tokyo 181-8588, Japan}

\begin{abstract}
Dense core collisions, previously regarded as minor in star formation, are proposed to play a significant role in structure formation around protostellar envelopes and binary formation.
Using archival data of nearby star-forming regions, we determine the frequencies of core collisions.
Our calculations reveal that a typical core is likely to undergo multiple interactions with other cores throughout its lifetime. 
To further investigate the core collision process, we employ adaptive mesh refinement hydrodynamic simulations with sink particles.
Our simulations demonstrate that following the formation of a protostar within a gravitationally-unstable core, the merging core's accreting gas gives rise to a rotationally-supported circumstellar disk. Meanwhile, the region compressed by the shock between the cores develops into asymmetric arms that connect with the disk. Gas along these arms tends to migrate inward, ultimately falling toward the protostar.
One of the arms, a remnant of the shock-compressed region, dominates over the second core gas, potentially exhibiting a distinct chemical composition. This is consistent with recent findings of large-scale streamers around protostars.
Additionally, we found that collisions with velocities of $\sim$ 1.5 km s$^{-1}$ result in the formation of a binary system, as evidenced by the emergence of a sink particle within the dense section of the shocked layer.
Overall, dense core collisions are highlighted as a critical process in creating $10^3$ au-scale streamers around protostellar systems and binary stars.
\end{abstract}


\section{Introduction}
\label{sec:intro}

The standard scenario of star formation proposes that once a dense core forms from a parent molecular cloud, it undergoes gravitational collapse to form a star or small stellar system such as binary or multiple stars \citep{shu87}.  
Since molecular clouds are highly turbulent, a core formed has small angular momentum prior to its gravitational contraction. Consequently, the contraction process leads to the formation of a protostar and its surrounding rotating disk, which typically exhibit nearly axisymmetric structures. However, recent high-resolution millimeter and submilliemeter observations have unveiled significant asymmetries in circumstellar structures, such as filaments, streamers, and spirals, which appear to be connected to the circumstellar disks \citep{tokuda14,pineda20,chen21,sanhueza21,valdivia22,hsieh23}.
These structures often extend over substantial distances from the central protostars, spanning $10^3-10^4$ au, and sometimes exhibit linear velocity gradients along the streamers \citep{sanhueza21,valdivia22,hsieh23}. These velocity structures can be interpreted as inflow motions towards the protostar and its disk.

On a large scale, several theoretical and observational studies have indicated that dense cores have the potential to accumulate additional mass from their surrounding environments. Extensive numerical simulations of cluster formation have supported this notion, revealing that cores can acquire a significant amount of gas from their surroundings \citep[see also][]{bonnell06,wang10,pelkonen20,kuznetsova22}.
From the comparison between the core mass function (CMF) and stellar initial mass function (IMF) in the Orion Nebular Cluster region 
(using the CARMA-NRO Orion Survey data \citep{kong18}), 
\citet{takemura21a} discovered that the estimated mass of typical protostellar cores was approximately double that of prestellar cores in Orion A. This finding suggests that the current CMF can potentially replicate the IMF if cores experience mass growth through the accretion of surrounding gas \citep{takemura21a}. Moreover, several observational studies have reported evidence supporting core growth and accretion \citep{contreras18,kong21}.

An additional mechanism contributing to core growth is the merging of cores.
Barnard 68 serves as an excellent observational example of core growth resulting from core collisions \citep{burkert09}.
Interestingly, dense Bok globules, which typically exist in relative isolation, often exhibit the presence of subcores in different evolutionary stages.
This phenomenon is well demonstrated in the Herschel EPoS (Early Phase of Star Formation program) target, where 11 out of 12 targets analyzed by \citet{launhardt13} contain secondary, less-dense compact structures near the densest cores. These structures are typically elongated rather than spherical in shape. For example, starless systems like CB4, CB26, CB27, and B68 exhibit subcores, while protostellar systems such as CB6, CB17, CB27, BHR12, CB68, CB130, CB224, and CB230 also possess subcores (refer to Figures 12 through 23 in \citet{launhardt13}). B335 stands as the only target without apparent subcores, but it features an extended envelope with several arm-like structures, likely resulting from recent core collisions or the accretion of ambient gas.
Given the isolated nature of Bok globules, the primary core and adjacent subcores gravitationally influence one another and are susceptible to merging. An illustrative case is CB17, a extensively-studied Bok globule, which contains a starless core (SMM1) and a Class I source separated by about 5000 au \citep{schmalzl14}. These two cores appear to be connected by a bridge in the position-velocity diagrams of N$_2$H$^+$ \citep{schmalzl14} and NH$_3$ \citep{spear21}, with a measured velocity difference of 0.3 km s$^{-1}$ along the line-of-sight. Such a bridge structure is likely the result of a collision \citep[e.g.,][]{haworth15,wu20}.

The phenomenon of core collision and merging has been shown to generate asymmetric structures \citep{kuffmeier21,kinoshita22,
hanawa22}.
Recent simulations have further demonstrated that core collisions can explain several characteristics that are challenging to account for solely with the isolated core collapse model. Particularly, core collisions can create protostellar systems with misaligned disks and even counter-rotating disks \citep{kuffmeier21,kuznetsova22}. Hence, it is crucial to investigate the processes and resulting structures that emerge during core collisions.

\citet{kinoshita22} conducted magnetohydrodynamic simulations to study the collision of two equally massive, gravitationally-stable prestellar cores. Their work illustrated that core collisions lead to the formation of single or multiple protostars. In this study, we extend the model proposed by \citet{kinoshita22} to examine the collision between cores of unequal mass, specifically focusing on the collision between a gravitationally unstable core and a stable core. For simplicity, we neglect the effects of magnetic fields and initial core rotation in our present analysis.
It is worth noting that filament collisions have been observed to influence structure formation in molecular clouds \citep{nakamura14,frau15}. 
Such filament collisions are likely to trigger core collisions in crowded regions.

The paper is organized as follows. 
Before showing the results of the numerical simulations, we evaluate how frequently the core collision happens in nearby star-forming regions in Section \ref{sec:frequency}.
In Section \ref{sec:methods}, we briefly describe our numerical methods used in the simulations. 
In Section \ref{sec:results}, we present our numerical results.
Then, we discuss some important features of the core collision process in Section \ref{sec:discussion}.
Finally, we summarize our results in Section \ref{sec:summary}.

\section{Collision frequency in nearby star-forming regions} 
\label{sec:frequency}

Before conducing the numerical simulations of the core collisions, we attempt to roughly estimate the core collision frequencies in observed molecular clouds. 
Here, we consider the two representative situations.
One is the case in which the cores are moving randomly in molecular clouds. The second is that the cores form by gravitational fragmentation.

\subsection{randomly-moving cores}
For the first case, we follow the 
same method as \citet{kinoshita22}'s.
We use the observational data of nearby star-forming regions such as Orion A (OMC-1/2/3, OMC-4/5, L1641N, L1641C), $\rho$ Oph, 
Serpens South, Lupus I, and AGAL014.492-00.139.
These regions have very different cloud environments.
Orion A is the nearest giant molecular cloud, containing high-mass/intermediate-mass star-forming region (OMC-1/2/3), low-mass star cluster-forming regions (L1641N, L1641C), and distributed star formation regions (OMC-4/5).  $\rho$ Oph and Serpens South are nearby low-mass star cluster-forming regions.
Lupus I is the nearby distributed star formation region.  
AGAL014.492-00.139 is the densest molecular clump in the very early phase of high-mass star formation \citep{morii21}.

Here, we estimate the collision frequency in each star-forming region by considering that identical spherical cores with mean radius
$\left<R_{\rm c}\right>$ and their mean velocity of $\left< V_c\right>$ are distributed uniformly in the 3D space with a core number density of $N_c$.  
For such a situation, the collision timescale can be estimated as
\begin{eqnarray}
\tau_{\rm col}&=&{\lambda \over \left< V_c\right>}
={1\over 4 \pi R_c^2 N_c \left< V_c\right>}  \\
&=&
 0.311 \ {\rm Myr} \left({R_c \over 0.05 {\rm pc}}\right) ^{-2} \left({N_c \over 10^2 {\rm pc}^{-3}}\right)^{-1} \left({\left<V_c\right> \over 1 {\rm km} \ {\rm s}^{-1}}\right)^{-1}   \ ,
\end{eqnarray}
where we neglect several factors such as magnetic field and gravity. 
Then, we define the core lifetime $\tau_{\rm life}$ as the sum of the prestellar core lifetime and the protostellar (=Class 0/I) lifetime.
Molecular line observations of nearby star-forming regions suggest that the majority of the cores are gravitationally-unbound \citep{maruta10,takemura23}.  
The statistical studies of the prestellar lifetimes suggest that the typical prestellar lifetime ranges from 1 $t_{\rm ff}$ to 10  $t_{\rm ff}$.
Therefore, we choose $5 t_{\rm ff}$ \citep{beichman86,jessop00,takemura23} as a representative value.
The core free-fall time is defined as
\begin{eqnarray}
t_{\rm ff} &=&  \left({3 \pi \over 32 G \rho_c } \right)^{1/2}    \\
&=& 0.347 \ {\rm Myr} \left({n_c \over 10^4 {\rm cm}^{-3}}\right) ^{-1/2}
\end{eqnarray}
where $\rho_c$ and $n_c$ are the volume density and number density of the core, respectively, and we assume the mean molecular weight of 2.3.

The lifetime of the Class 0/I objects is estimated to be 0.5 Myr \citep{evans09}.
The core lifetime is then given by
\begin{equation}
\tau_{\rm life} = 5 t_{\rm ff} + 0.5 \ {\rm Myr}
\end{equation}
Thus, for the core density of $10^4$ cm$^{-3}$ to $10^5$ cm$^{-3}$, the core lifetime is calculated to be 
about 1 Myr.
The collision frequency is defined as the ratio of the core lifetime to the collision time.

\begin{table*}
\tiny
 \centering
  \caption{The collision time scale}
  \begin{tabular}{lccccccccccc}
 \hline \hline 
Region & $r_c$ & n & $\Delta V_{\rm 1D}$ & $\left<V_{\rm c}\right>/ c_s$ & $N$ & Volume & $\tau_{\rm col}$  & $t_{\rm ff}$ & $\tau_{\rm life}$ 
& $\tau_{\rm life}/ \tau_{\rm col}$ & Reference\\
 & (pc) & ($\times 10^4$ cm$^{-3}$) & (km s$^{-1}$) &  & & (pc$^3$) & (Myr) & (Myr)  & (Myr)  & \\
 \hline
OMC-1/2/3 & 0.066 & 5.94  & 2.18 & 20.1--14.2  & 716 &  a cylinder with $(r,z)=$ (1.7, 5) pc  & 0.300 & 0.142 & 1.21 & 4.1 & 1 \\
OMC-4/5 & 0.077 & 1.53  & 1.10 & 10.1--7.14 &605 & a cylinder with $(r,z)=$ (1.8,5) pc  & 0.579 & 0.280 & 1.90 & 3.3 & 1 \\
 L1641N & 0.082 & 1.13  & 1.30 & 12.0--8.48 &726 &  a cylinder with $(r,z)=$ (2,5) pc & 0.44 & 0.326 & 2.13& 4.8 & 1 \\
  L1641C & 0.082 & 1.85  & 0.99 & 9.12--6.45
  & 294 &  a cylinder with $(r,z)=$ (1.7,3.5) pc & 0.73 & 0.255 & 1.78 & 2.4 & 1 \\
$\rho$ Oph & 0.045 &12.5  &  0.57 & 4.90--3.46  & 68 & a sphere with $r=$ 0.3 pc & 0.065 & 0.098 & 0.99 & 15.3 & 2 \\
Serpens South & 0.0649 & 9.6 & 0.50 & 4.61--3.26& 70 & a cylinder with $(r,z)=$ (0.6, 1.2) pc & 0.41 & 0.112 & 1.06 & 2.6 & 3 \\
AGAL014.492-00.139 & 0.0201 & 137.9 & 1.17 & 4.4--6.2 & 28 & a sphere with $r=$ 0.2 pc & 0.11 & 0.03 & 0.65 & 5.7 & 4\\
Lupus I & 0.0233 & 1.50 & N/A (0.5) & N/A(4.90--3.46) & 25 & a sphere with $r=$ 0.5 pc & 3.5 & 1.33 & 1.92 & 0.55 & 5\\
\hline 
\end{tabular}
   \begin{tablenotes}
\item[ ] 
$\Delta V_{\rm 1D}$ is the one-dimensional core-to-core velocity dispersion along the line of sight. We note that the 1D core-to-core velocity dispersions are essentially equal to the one predicted from the linewidth-size relation.
$\left<V_c\right>$ is calculated as $\sqrt{3} \times \Delta V_{\rm 1D}$, assuming the isotropic motion.
 $\tau_{\rm life}$ is defined as 5$t_{\rm ff}$ (prestellar lifetime) + 0.5 Myr (Class 0/I).
 \item[]
 For OMC-1/2/3, OMC-4/5, L1641N, L1641C, and Serpens South, the cores are assumed to be distributed in a cylindrical space, whereas for $\rho$ Oph, we assume the cubic geometry.
 \item[]
 $c_s$  ($=0.266 \ {\rm km \ s^{-1}} \sqrt{T/15 \ {\rm K}}$) is the isothermal sound speed 
 and we assume $T=10-20 $K.
 \item[] For Lupus I, we choose the most crowded part and adopt the $\Delta V_{1D}$ of $\rho$ Oph, as its typical value. 
 For AGAL014.492-00.139, we adopt $T=15$ K to estimate the core density.
 The total core number adopted is the sum of prestellar and protostellar core numbers.
 \item[] For references, 1: \citet{takemura23}, 2: \citet{maruta10}, 3:\citet{tanaka13}, 4: \citet{redaelli22}, 5: \citet{benedettini18}
 \end{tablenotes}
  \label{tab:obs}
\end{table*}

Below, we used all the bound and unbound cores to estimate the collision frequencies since the Orion core survey by \citet{takemura21a} suggests that all the cores should contribute star formation to match the estimated current star formation rate. A brief discussion is as follows.
In cluster-forming regions, the supersonic turbulence is maintained by the momentum injection due to the stellar feedback 
\citep{li06,nakamura07,krumholz14,offner17}. In such an environment, a significant fraction of the cores can be pressure-confined \citep{nakamura14b}.  In the Orion Nebular Cluster (ONC) region, we can roughly evaluate the current star formation rate to be SFR$_{\rm current}\approx $ 1000 stars /2 Myr $\approx$ 500 stars/Myr.
On the other hand, bound (subvirial) cores are found to be $\approx$ 100 cores, out of 700 cores identified in the ONC region.  The typical lifetime of the starless cores is estimated to be 
about 5 times the free-fall time which is about 2 Myr.   Thus, if only the bound cores form stars, the expected star formation rate is only SFR$_{\rm future} \approx$ 50 stars / Myr.  To match the current and expected SFR (SFR$_{\rm current} \approx $ SFR$_{\rm future}$), 
each core should form 10 stars, which we think too many and therefore unlikely.
If all the identified cores form stars, the expected star formation rate is SFR$_{\rm future} \approx $ 300 stars/Myr, which is comparable to the current star formation rate. 
When all the cores creates binaries, SFR$_{\rm future} \approx $ 600 stars/Myr, almost equal to SFR$_{\rm current}$.
Therefore, we consider that both bound and unbound cores should contribute to the future star formation.

Table \ref{tab:obs} summarizes the adopted physical quantities and estimated collision frequencies for several nearby star-forming regions.  
For Lupus I, the velocity dispersion is not available and we simply used the same value as $\rho$ Oph, the minimum value among the regions listed here. For Lupus I, we also choose only a dense part of 1 pc$^3$.
For all the clouds, the typical core has a radius of about 0.05 pc with a density of $10^{4-5}$ cm$^{-3}$.
The collision velocity is estimated to be 0.4 km s$^{-1}$ -- 3.5 km s$^{-1}$.  Therefore, it is
about 3--20 times the sound speed with $T=10$--15 K.
We note that the collision frequency and time depend on the assumed volume that cores are distributed. For simplicity, we adopted a sphere or cylinder that covers the entire area. In this sense, the collision frequencies calculated here are likely to be close to the lower limits. 

The 11th column of Table \ref{tab:obs} gives the calculated collision frequencies.
In clustered environments such as $\rho$ Oph OMC-1/2/3, and AGAL014.492-00.139, a typical core is expected to experience several collisions in its lifetime. Even in the prestellar phase ($\approx 5t_{\rm ff}$), the collision frequency is of the order of unity.
For more distributed environments such as OMC-4/5, the collision frequency is estimated to be a few. 
In Lupus I, although the core velocity dispersion is not available, the collision frequency is somewhat lower than unity. However, the frequency of 0.55 suggests that several (or $\sim$ 50\%) cores in this area should experience single collision. 
In other words, a typical core in nearby star-forming region experiences at least one collision with the other core in its lifetime, and in the clustered environments, multiple collision is likely to occur.

\citet{offner22} investigated time evolution of cores formed in turbulent molecular clouds, and estimated that about 37\% and 32 \% of cores disperse and merge with each other, respectively. We consider their core dispersed rate is an upper limit estimated from the simulations with no stellar feedback which  maintains the supersonic turbulence.
The strong turbulence should be important to confine the cores by external pressure in real molecular clouds.
Therefore, we infer that the core merging rate should be higher than the above value in real molecular clouds. 

\subsection{Gravitational  Fragmentation}
Above, we considered that the cores are randomly-moving in a molecular cloud. In reality, the cores may form by gravitational fragmentation. In that case, the separations between two adjacent cores are of the order of the thermal Jeans length
(e.g., Ishihara et al. 2023), 
\begin{eqnarray}
L_{\mathrm{J}} &=& c_s \left( \frac{\pi}{G \rho_c} \right)^{\frac{1}{2}}  \ , \\
    &=& 0.2 \ {\rm pc} \left( {T \over 10~{\rm K}} \right)^{1/2} \left( {n_c \over 10^4~{\rm cm^{-3}}} \right)^{-1/2}  \ , 
    \label{eq:termal_JL}
\end{eqnarray}
where $c_s = (k_BT/\mu m_H)^{\frac{1}{2}}$ is the isothermal sound speed, 
$G$ is the gravitational constant, and $k_B$ is the Boltzmann constant, $\mu$ (=2.3) is the mean molecular weight, and $m_H$ is the hydrogen mass.
Here we consider that the two cores with a separation of the Jeans length  are approaching with a velocity of $V_{\rm rel}$. 
The collision time is then estimated to be
\begin{eqnarray}
\tau_{\rm col}^{\rm frag} &=& {L_{\rm J} \over V_{\rm rel}} = 
3.3 {c_s \over V_{\rm rel}} t_{\rm ff} \\ &=& 
0.2 {\rm Myr}  \left( {T \over 10~{\rm K}} \right)^{1/2} \left( {n_c \over 10^4~{\rm cm^{-3}}} \right)^{-1/2}  \\
& & \times \left ({V_{\rm rel} \over 1 \ {\rm km \ s^{-1}}}\right)^{-1} \ ,
\end{eqnarray}
Then, the collision frequency is calculated as
\begin{equation}
    \tau_{\rm life}/\tau_{\rm col}^{\rm frag} = 1.5 {V_{\rm rel} \over c_s} \left(1 + {0.1 \ {\rm Myr} \over t_{\rm ff}} \right)
\end{equation}
This frequency is larger than unity for the cores approaching with $V_{\rm rel} \sim c_s$.
The observed velocity dispersion ranges from 20 to 3 times the sound speed in the above table, 
and therefore the cores created by gravitational fragmentation are prone to collide with adjacent cores.

In summary, we conclude that the core collision is likely to happen frequently for both clustered and distributed environments. 
In the following sections, we investigate the core collision by using the numerical hydrodynamic simulations.

\section{Numerical Methods}
\label{sec:methods}

\begin{figure}[htbp]
    \plotone{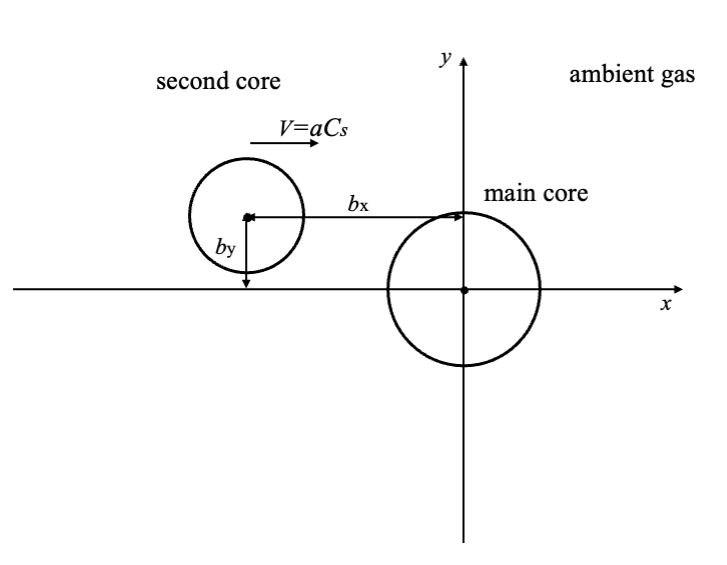}
    \caption{Schematic diagram of the initial condition.}
    \label{fig:model}
\end{figure}

\subsection{Initial Model}

\begin{table}
 \centering
  \caption{Initial Core Properties}
  \begin{tabular}{cccc}
 \hline \hline 
  & Main core & Second core & Ambient gas  \\
 \hline
Radius ($R_{\rm BE}$) & 1.0  & 0.75 &  $\cdots$\\
Radius (pc) & 0.074  & 0.056 &  $\cdots$\\
T (K)  & 10  & 10 & 500\\
Mass ($M_\odot$) & 2.2 & 1.0 & $\cdots$ \\
$\rho_c$ (g cm$^{-3}$) & 5.01$\times 10^{-19}$ & 3.34 $\times 10^{-19}$  & 1.00 $\times 10^{-20}$  \\
\hline 
\end{tabular}
  \label{tab:model1}
   \begin{tablenotes}
 \item[ ] 
 $\rho_c$ denotes the core central density, and is ambient uniform density for ambient gas.
 \end{tablenotes}
\end{table}
We consider two isothermal spherical cores which collide with each other.
Figure \ref{fig:model} shows the schematic diagram of the initial condition.
A main core is centered at the computation box and is stationary at $t=0$. We adopt the Cartesian coordinate system, ($x,y,z)$. 
The computation box is set from $(x,y,z)=(-0.37 {\rm pc}, -0.37{\rm pc},-0.37{\rm pc})$ to $(+0.37 {\rm pc}, +0.37{\rm pc},+0.37{\rm pc})$.
A less-massive core, referred to the second core, is placed at the position where the core center is equal to $(x,y,z)= (-b_x, 0, b_z)$. 
The value of $b_z$ is essentially  the collision impact parameter. 
The second core has a velocity of $v= a c_s$, where $c_s$ is the isothermal sound speed with the core temperature of $T=10$ K.  

This initial model is similar to that of \citet{kinoshita22} whose initial model was the collision of two identical stable cores. 
The core density profiles are similar to the hydrostatic equilibrium isothermal sphere, that is the Bonner-Ebert sphere \citep{bonnor56}. 
The density profile of a main core is set to be similar to the critical Bonner-Ebert sphere with a radius of $R_{\rm BE}$, but we multiplied a factor of 1.5 in the density to induce the gravitational collapse at the beginning.  
The second core is a stable Bonner-Ebert sphere with a radius of $0.75 R_{\rm BE}$. It is set to be in pressure equilibrium at the cloud surface with ambient uniform gas with the density of 1 $\times 10^{-20}$ g cm$^{-3}$ and temperature of 500 K. Its total mass is  1.0 M$_\odot$.  
The central density of the second core is 1.5 times smaller than that of the main core, 5.01 $\times  10^{-19}$ cm$^{-3}$. The main core is in nearly pressure equilibrium at the cloud surface where the density is $3\times 10^{-20}$ g cm$^{-3}$ and therefore the external pressure ($ P/k_B \simeq 1 \times 10^5$ cm$^{-3}$ K) somewhat larger than the cloud internal pressure ($nT\simeq 0.8\times 10^{5}$ cm$^{-3}$ K).  
The total mass is 2.2 M$_\odot$, about twice the second core.

Tables \ref{tab:model1} and \ref{tab:model2} summarize the initial parameters of the simulation. 
We adopted the physical quantities which are consistent with those of the nearby star-forming regions summarized in Table \ref{tab:obs}.

\begin{table}
 \centering
  \caption{Numerical Model}
  \begin{tabular}{lccc}
 \hline \hline 
Name & $V_{\rm col}/ c_s$ & $b_z/R_{\rm BE}$ & Note  \\
 \hline
A2B05 & 2 & 0.5 & slow \\
A4B05 & 4  & 0.5  &  slow \\
A5B05 & 5 & 0.5  &  fiducial model \\
A6B05 & 6  & 0.5  &  fast \\
A7B05 & 7  & 0.5  &  fast \\
A5B025 & 5  & 0.25  & small $b_z$\\
A5B075 & 5  & 0.75  & large $b_z$\\
\hline 
\end{tabular}
  \label{tab:model2}
\end{table}

\subsection{Numerical Code}

We use Enzo, a hydrodynamic adaptive mesh refinement (AMR) core. See \citet{bryan14} for the details of the numerical code.
The top-level root grid is set to $256^3$ with 5 additional levels of refinement. This corresponds to an effective resolution of $4096^3$.
For refinement conditions, we adopt the Jeans condition and slope condition.  For the former, we set the Jeans number to $n_J=$ 8. When the further refinement level is needed, a sink particle is created.  At the maximum level, the cell size is equal to about 37 au. 

Following \citet{kinoshita21} and \citet{kinoshita22}, we use a color variable, $C$. In our simulations, we assigned ``1'', ``-1'', and ``0'' to the main core, second core, and ambient gas.

\begin{figure}[htbp]
\includegraphics[width=\linewidth]{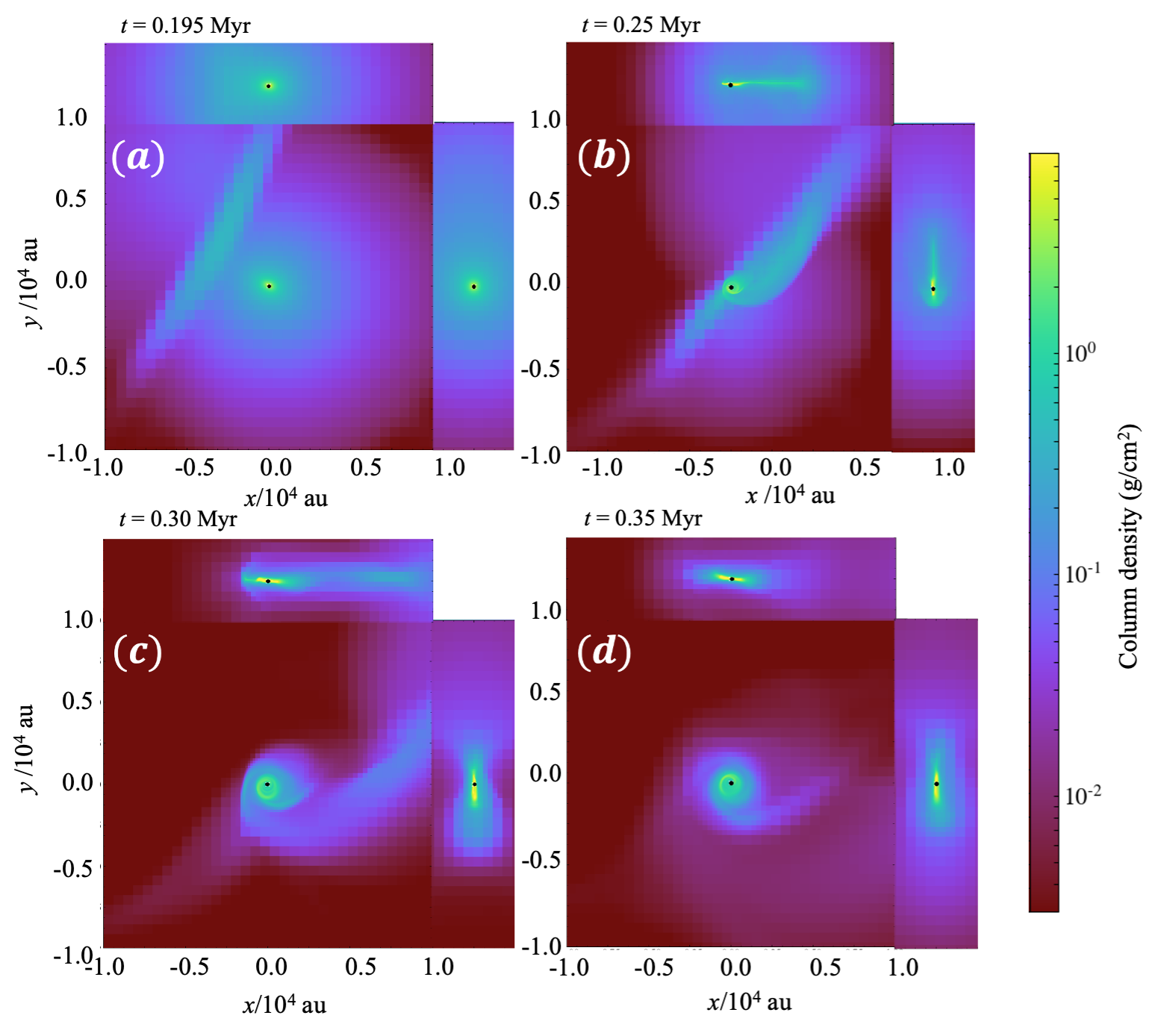}
    \caption{Column density distributions projected on
    the $x$-$y$ ({\it center}), $z$-$y$ ({\it right}), and $x$-$z$ ({\it upper}) planes at $t=$ 0.195, 0.25, 0.30, and 0.35 Myr.
    Only the central 20000 au $ \times$ 20000 au region is shown. The sink particle is indicated with the black dots.}
    \label{fig:modela}
\end{figure}

\section{Numerical Result}
\label{sec:results}

\subsection{Fiducial Model A5B05}

Here, we describe the evolution of model A5B05, in which the collision speed is set to $5c_s = 0.94$ km s$^{-1}$ and the impact parameter of $b_z= 0.5 R_{\rm BE}$ = 0.037 pc.
Figure \ref{fig:density} shows the column density distribution on the slices at $x-y$, $x-z$, and $y-z$ planes at $t=$ 0.195, 0.25, 0.30, and 0.35 Myr. 

Since the main core is gravitationally unstable, it starts gravitational contraction immediately. 
The contraction proceeds similarly to the single spherical core collapse, and the density increases mainly at the central dense part.
The density distribution tends to approach to the $r^{-2}$ profile in the central dense part.
Such a feature is essentially the same as that of the isothermal spherical collapse \citep{larson69,foster93,ogino99}.
When the central density becomes large, a sink particle is created at the center, and the sink particle grows in mass monotonically by accretion of ambient gas. 

The second core collides with the main core by the time of $t\sim 0.195$ Myr. By that time, the sink particle was formed at the center of the main core.
Then, the shock-compressed layer appears at the upper-left part of the main core (Figure \ref{fig:modela}a).
The layer extends to a length of about $10^4$ au from the center.
By the time when about a half of the main core mass goes into the sink particle ($t\sim 0.23$ Myr), 
the disklike structure developed around the sink particle (see Figure \ref{fig:modela}b), 
and its size gradually becomes larger. 
As the time goes by, the rotating motion becomes more prominent in the disk, and the disk grows in size.
The spiral structure appears around the sink particle. 
The compressed layer appears to be a curved arm connecting to the central circumstellar disk.  
Hereafter, we call a curved arm a streamer.
By $t=0.35$ Myr, the extended compressed layer fragments into two (Figure \ref{fig:modela}c). The fragmented layer near the sink particle coils around the central disk. 
Another fragment is moving away from the sink particle and going out of the right side of the panel (Figure \ref{fig:modela}d).

\begin{figure*}[htbp]
\includegraphics[angle=0,width=\linewidth]{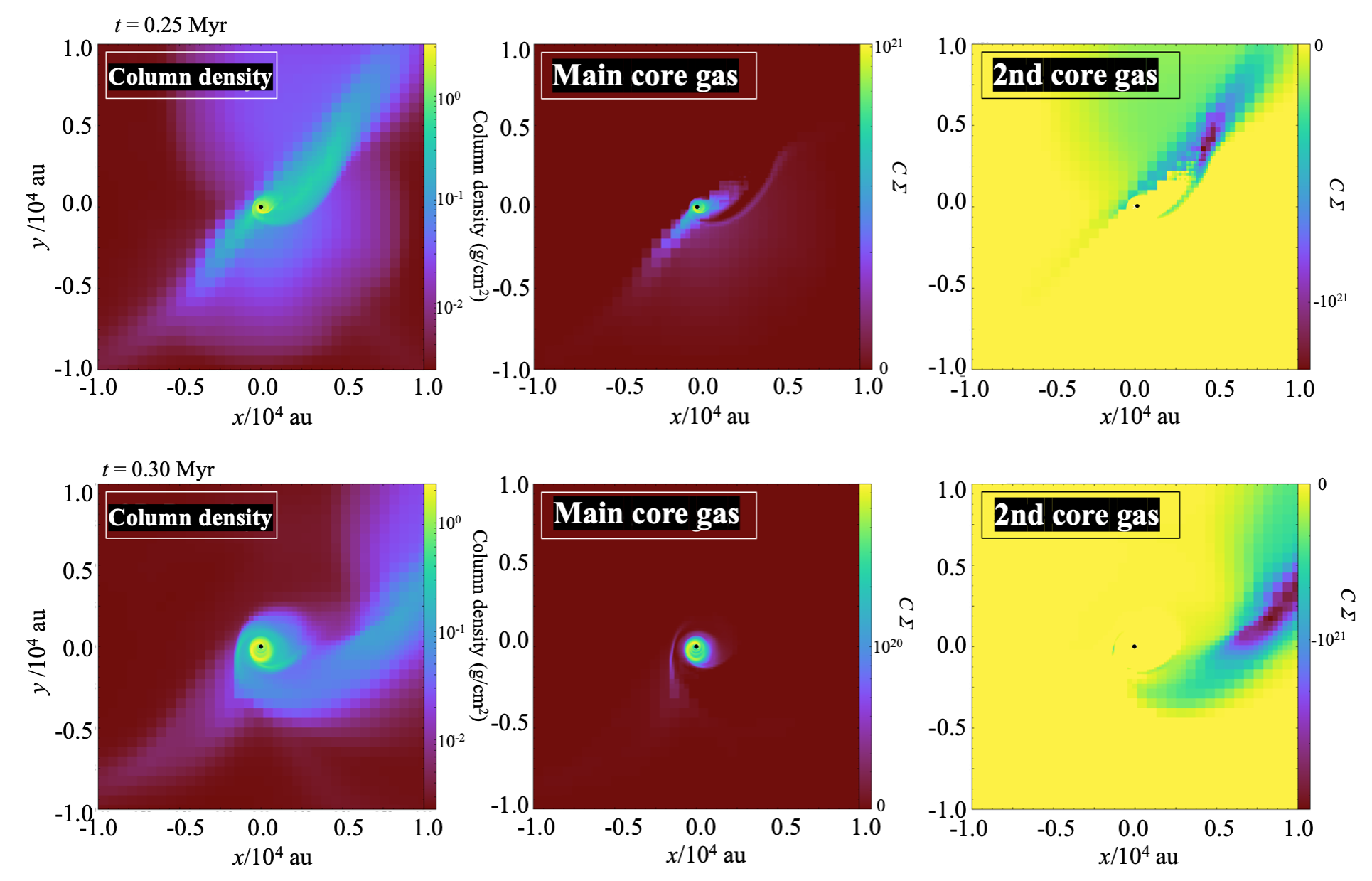}
    \caption{The spatial distributions of ({\it left}) the column density, ({\it center}) the density-weighted positive color values $C_0 \rho$, 
    and ({\it right}) the density-weighted negative color values at $t=$ 0.25 Myr and 0.30 Myr.  The back dots are the positions of the sink particle.
}
    \label{fig:color}
\end{figure*}

\begin{figure}[thbp]
\includegraphics[width=\linewidth]{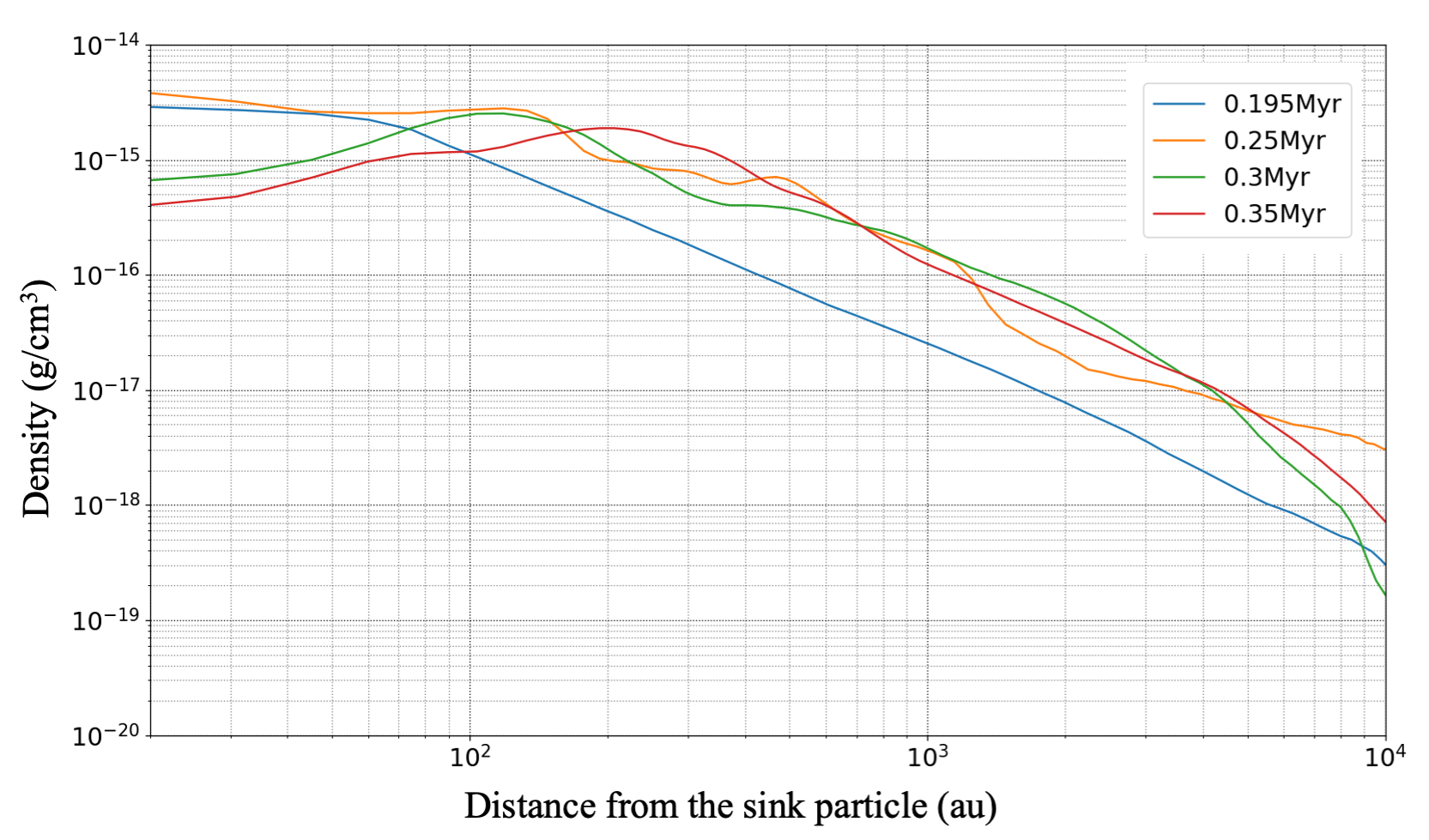}
    \caption{Circular-averaged column density  distributions projected on the $z=0$ plane, measured from the position of the sink particle, at the time of $t=0.195$, 0.25, 0.30, and 0.35 Myr.}
    \label{fig:density}
\end{figure}

In Figure \ref{fig:color}, we
show the distributions of the  density-weighted color values
at $t=0.195$ ({\it upper panels}) and 0.30 Myr ({\it lower panels}).
The areas with the positive and negative color values mean that the gas initially belonging to the main core and second core, respectively, is dominated. 
In this model, the central disk mainly consists of the main core and the streamer comes from the second core.

\begin{figure}[htbp]
\includegraphics[width=\linewidth]{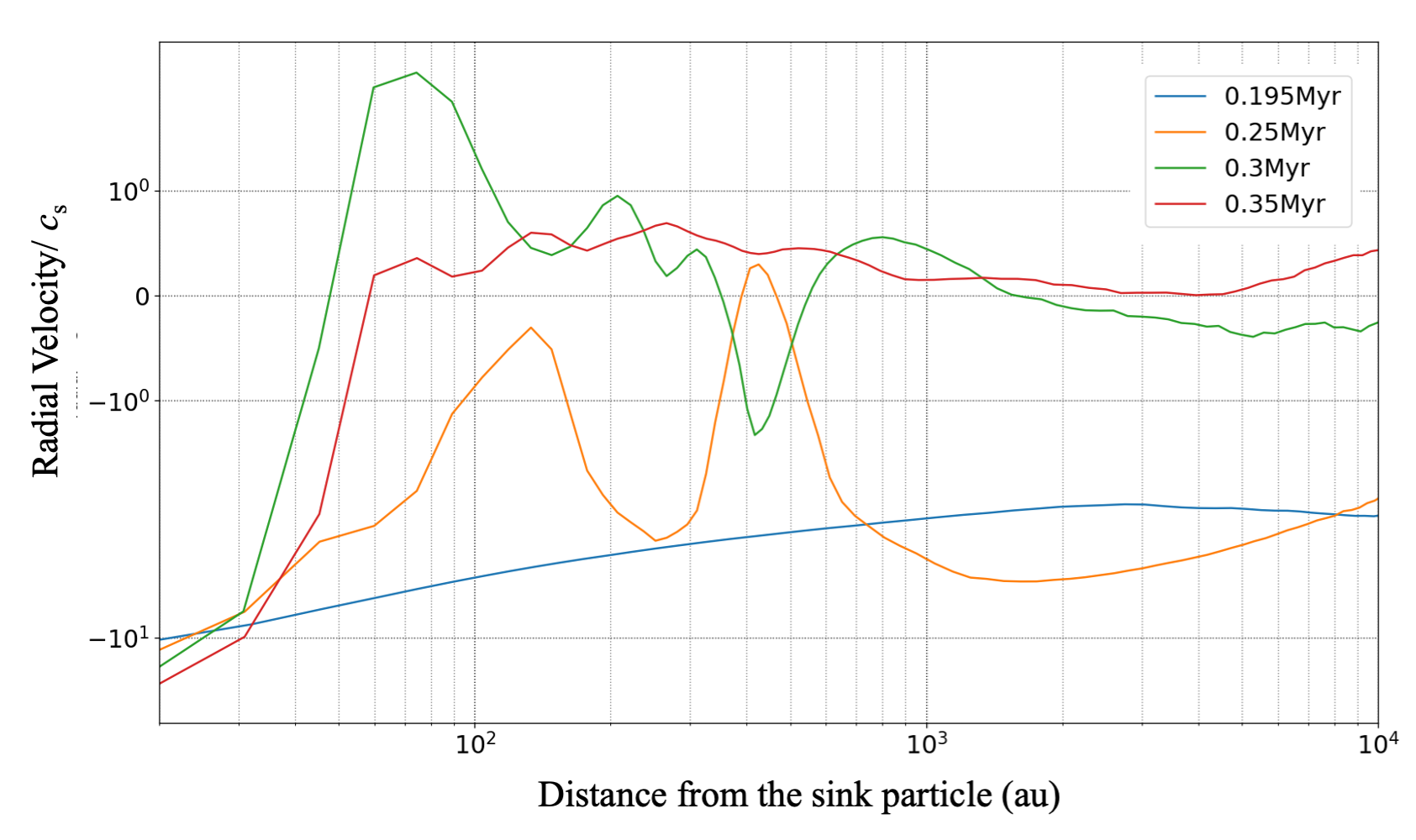}
    \caption{Circular-averaged radial velocity ($V_{\rm radial}$) distribution projected on the $z=0$ plane, measured from the sink particle at the time of 
    $t=$ 0.195, 0.25, 0.30, and 0.35 Myr.
    The horizontal and vertical axes are given in logarithmic and symmetric logarithmic scales, respectively.
    The solid and dashed lines indicate the distributions  before and after the sink particle formation, respectively.}
    \label{fig:radial}
\end{figure}
\begin{figure}[htbp]
\includegraphics[width=\linewidth]{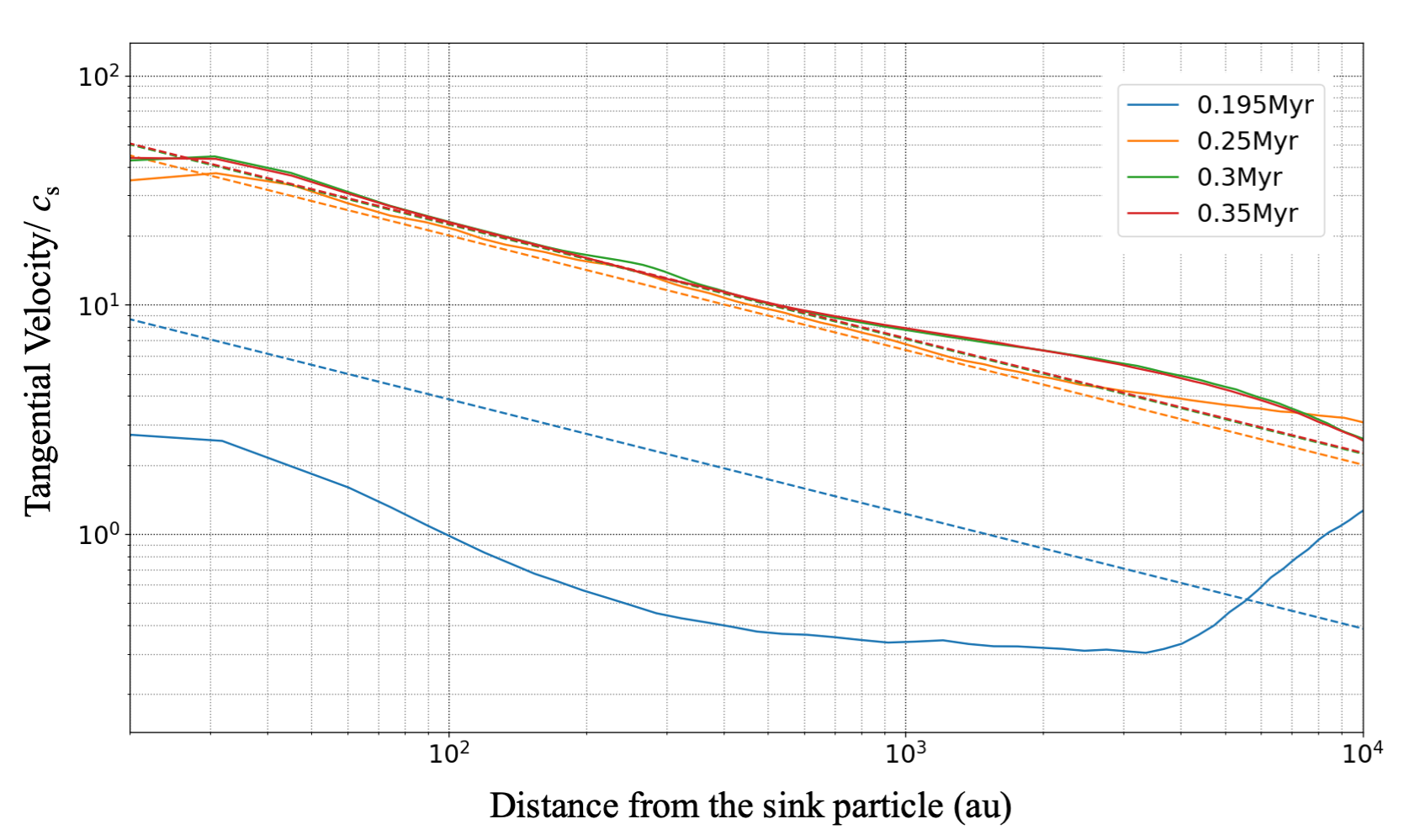}
    \caption{Circular-averaged tangential velocity ($V_{\rm tangential}$) distributions along the $z=0$ plane, measured from the position of the sink particle.
    The dotted lines represent the Keplerian rotation speed with the mass of the sink particle.}
    \label{fig:tangential}
\end{figure}

To see how the rotating disk grows, we show the column density, radial velocity, and tangential velocity distribution, circular-averaged from the sink particle, in Figures \ref{fig:density}, \ref{fig:radial}, and \ref{fig:tangential}, respectively.
Prior to the sink particle formation (By the time of 0.19 Myr), the $r^{-2}$ profile with a central flat part develops at the envelope of $r > 10^2$ au.
After the sink particle formation, the density distribution at the envelope have small undulation.
The column density distributions take peaks at around a few $\times 10^2$ au. Such a ring-like structure
is vaguely seen in the lower panels of Figure \ref{fig:density}.

The radial infall velocity increases with time toward the center of the main core, and after the sink particle formation, it takes a maximum at about 10 $c_s$ near the sink particle and declines toward the envelope. 
In other words, the radial infall is still significant at the later stages.
On the other hand, the tangential velocity is essentially zero before the sink particle formation, since the initial main core 
did not have any angular momentum.
After the sink particle formation, it gradually increases with time, and almost reaches the Keplarian rotation with the central sink mass, $V= \sqrt{GM_{\rm sink}/r}$, that is shown in the dashed lines in Figure \ref{fig:tangential}.

\begin{figure*}[thbp]
\includegraphics[angle=0,width=\linewidth]{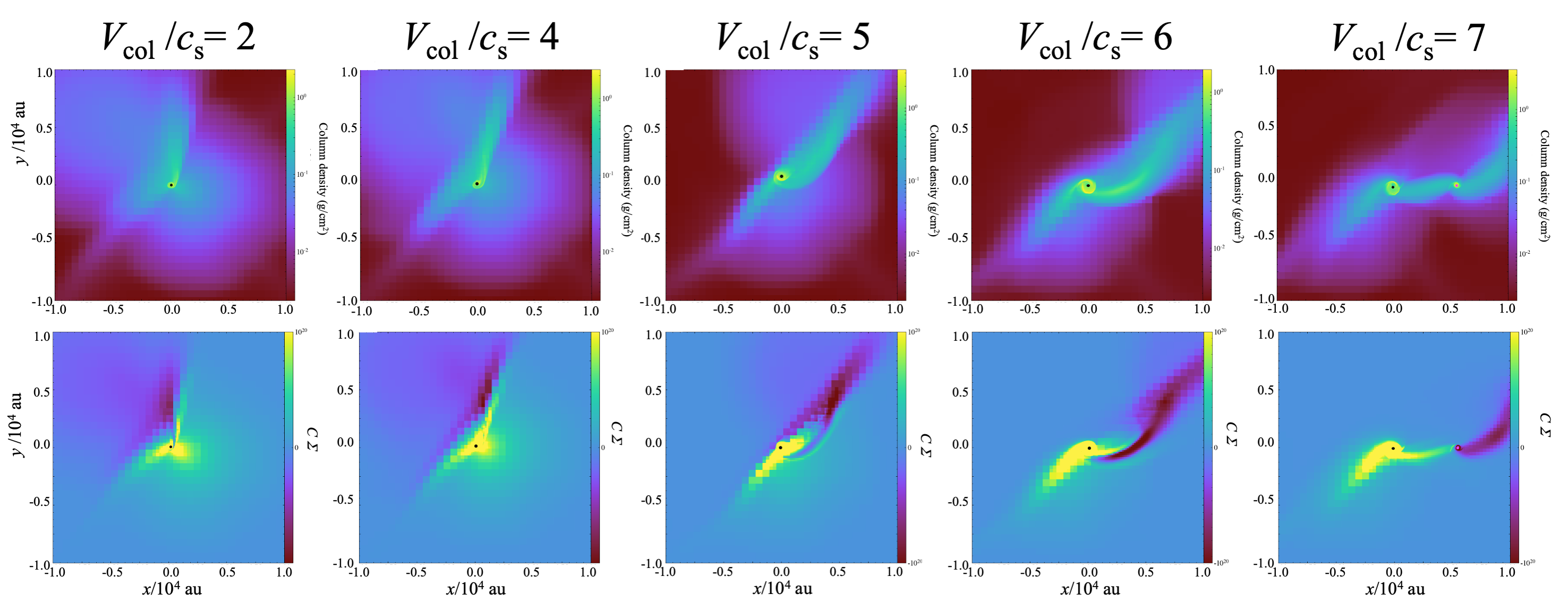}
    \caption{Column density ({\it upper}) and density-weighted color field ({\it lower} distributions in the $x-y$ plane for the models with different collision speeds
    at the evolution time of 0.25 Myr.
    From left to right, $V_{\rm col}$ = 2, 4, 5, 6, and 7 $c_s$.
    The black dots are the positions of the sink particles.
   For the fastest collision, the red dots are the positions of the second sink particle generated in the compressed layer.} 
    \label{fig:speed1}
\end{figure*}
\begin{figure*}[htbp]
\includegraphics[angle=0,width=\linewidth]{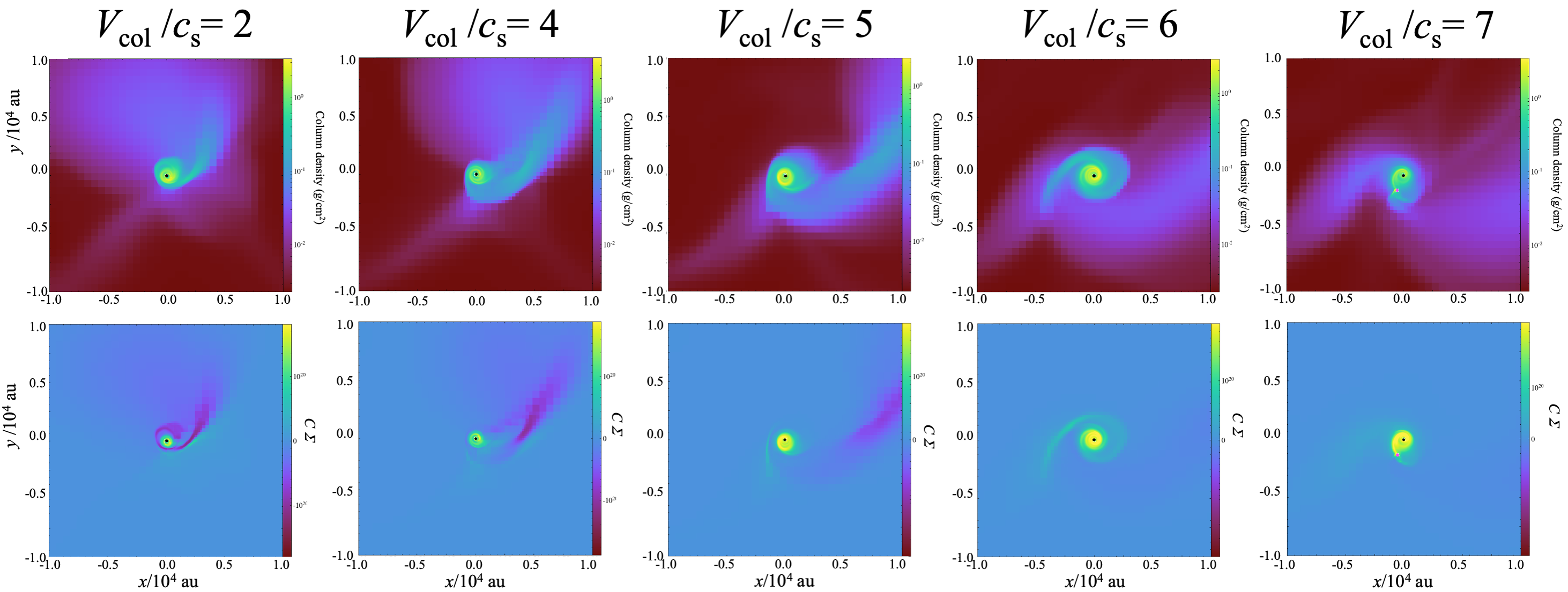}
    \caption{Same as that of Figure \ref{fig:speed1} but for $t=0.30$ Myr.}
    \label{fig:speed2}
\end{figure*}

\subsection{Dependence on the collision speed}
\label{subsec:speed}

Here, we examine how the collision speed influences the disk and streamer formation.
In Figures \ref{fig:speed1} and \ref{fig:speed2}, we present the column density distribution of the central region at $t=0.25$ and 0.30 Myr for the models with different collision speeds.

When the collision speed is small, 
the shock compression is weak and slow. Therefore, the gas of the second core falls around the sink particle or the protostar, after the main core contraction is almost completed.
Therefore, almost all the circumstellar disk gas originates from the second core gas.
On the other hand, as the collision speed is fast, the shock compressed layer becomes more prominent and it transfers more orbital angular momentum of the second core into the circumstellar gas and thus a larger rotating envelope tends to develop around the central protostar. 
The streamer is also created around the protostar, but it prone to be torn away from the circumstellar gas. 

For the model with the fastest collision speed ($7 c_s$), another protostar is formed in the shock-compressed layer, and eventually a binary system is created (Figures \ref{fig:speed1} and \ref{fig:speed2}).
The second star moves toward the primary star, and the separation becomes smaller (the leftmost panels of Figure \ref{fig:speed2}). In this simulation, we did not follow the further evolution of the binary, and thus the second star might eventually merge to the primary star.

\begin{figure}[htbp]
\includegraphics[angle=0,width=\linewidth]{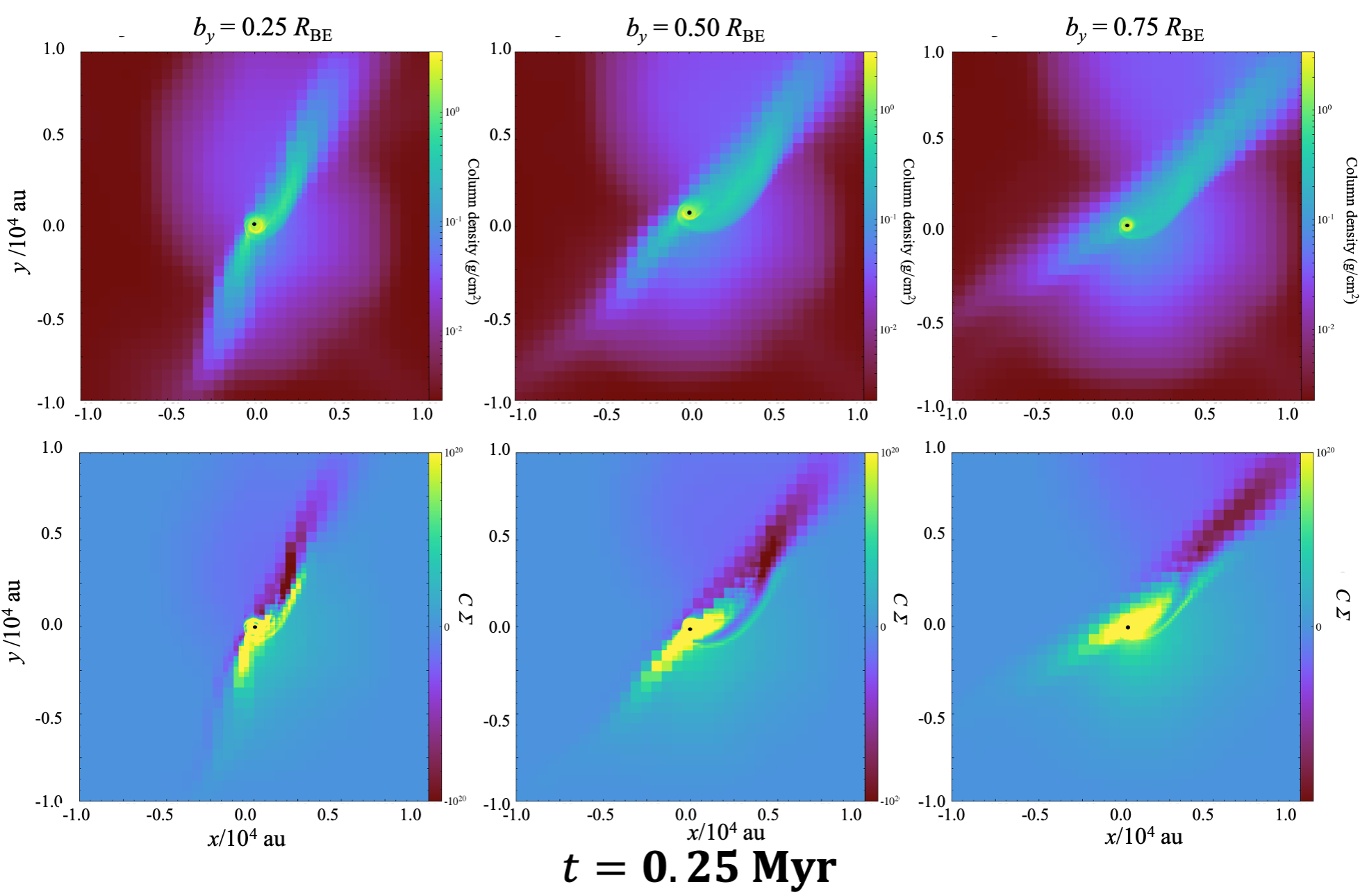}
    \caption{Same as Figure \ref{fig:speed1} but for the three models with different impact parameters.}
    \label{fig:impact1}
\end{figure}

\begin{figure}[htbp]
\includegraphics[angle=0,width=\linewidth]{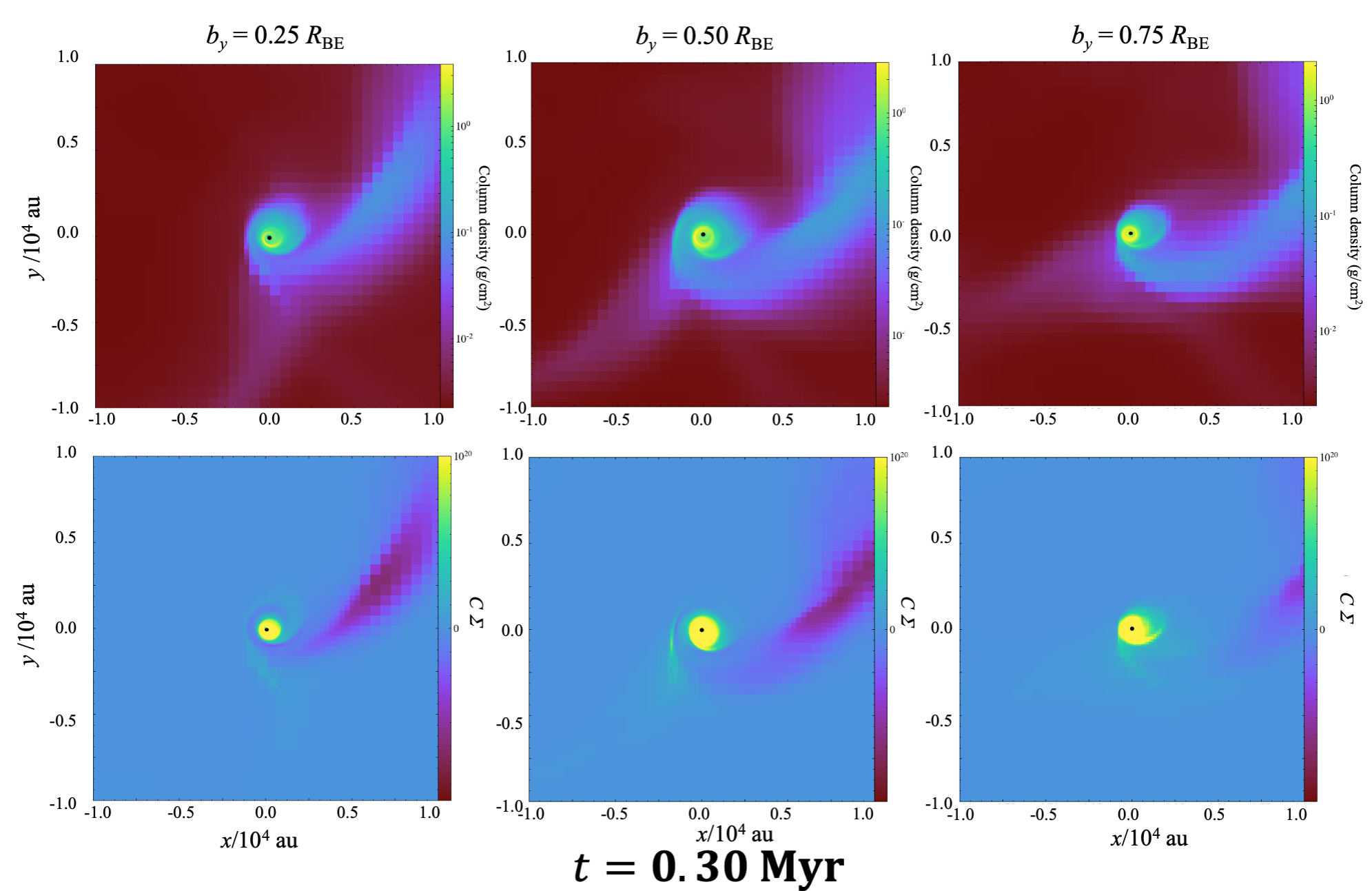}
    \caption{Same as Figure \ref{fig:speed2} but for the three models with different impact parameters.}
    \label{fig:impact2}
\end{figure}

\subsection{Dependence on the impact parameters}

The impact parameter is also important to 
influence the evolution of the core collision process.
The smaller impact parameter is expected to lead to the formation of denser, more massive compressed layers at the formation epoch of the sink particle since more gas is compressed by the collision.  In Figures \ref{fig:impact1} and \ref{fig:impact2}, we compare the column density distributions at $t=0.25$ and 0.30 Myr, respectively, for three models with different impact parameters $b_z=0.25$, 0.5, and 0.75.
For all the models, the collision speed is fixed to 5 $C_s$.
The initial cores, both the main and second cores, do not have rotation. Therefore, all the angular momentum originates from the orbital angular momentum of the second core. 
The orbital angular momentum of the second core is roughly proportional to the impact parameter and thus the impact parameter influences the formation and evolution of the rotationally-supported disk significantly.  In addition, the larger impact parameter leads to more subtle (weak) collision since the shock-compressed parts are less dense envelopes.
Therefore, among the models shown in Figures \ref{fig:impact1} and \ref{fig:impact2}, the disk size appears to be smallest for the A5B025 model with smallest impact parameter $b_z=0.25$.
This feature can better recognize for the density-weighted color field distribution. The central yellow part, which mainly consists of the main core gas, are larger for larger impact parameters at $t=0.30$ Myr.
At the later stage, the one-arm structure becomes more prominent for all the models.

\section{Discussion}
\label{sec:discussion}

\subsection{Observed Streamers and Spirals}

In Section \ref{sec:frequency}, we roughly estimated the core collision frequency in several star-forming regions. Our estimation indicates that  in clustered environments  a typical core is likely to collide with the other core a few times in its lifetime (prestellar and protostellar phases), 
Even in isolated environment, the collision is likely to frequently happen in its lifetime.
Therefore, the core collision is likely to be an important process in the structure process around the protostars. 
In fact, most of the EPoS targets, Bok globules, have less dense subcores around the densest cores.

One of the important roles of the core collision is that the orbital angular momentum of the merging core is used for the formation of circumstellar disk.
Using the numerical simulations, 
we have demonstrated that core collision can induce the formation of a protostar and rotating disk even in the absence of the initial core rotation. In addition, the disk is often associated with curved arms or streamer.  The collision model preferably creates one prominent arm or streamer.

Such asymmetric structures around the protostars, streamers and spirals, are often detected in both low-mass \citep{tokuda14,pineda20,hsieh23} and high-mass \citep{chen21,sanhueza21} protostars. 
Some streamers have different chemical compositions from the central protostellar system \citep{tokuda18,pineda20,chen21}. 
According to our simulations, one of the arms mainly contains the gas from the second core, whereas the other does the main core.  
Therefore, the chemical composition of the streamer can be different from that of the central protostellar system. 
Particularly, the gas from the second core is shock-compressed, and therefore, some shock tracer molecules such as SO are likely to become abundant.
In fact, the streamer observed by \citet{pineda20} contains the chemically-young molecular species such as CCS.
The core-to-core velocity dispersion measured from the line-width-size relation is typically several or 10 times as large as the sound speed. The typical core collision may increase due to dynamic compression the gas temperature at most $< 10^2$ K. Thus, the streamers formed by the core collision may be abundant in SO.
The steamers so formed also have inflow along them, feeding the gas of the second core to the central protostar. At the same time, the outer parts of the streamers torn away from the central part due to the bulk motion. They do not contribute to the accretion onto the central protostar, but makes the velocity structure complicated.

\begin{figure}[htbp]
\includegraphics[width=\columnwidth]{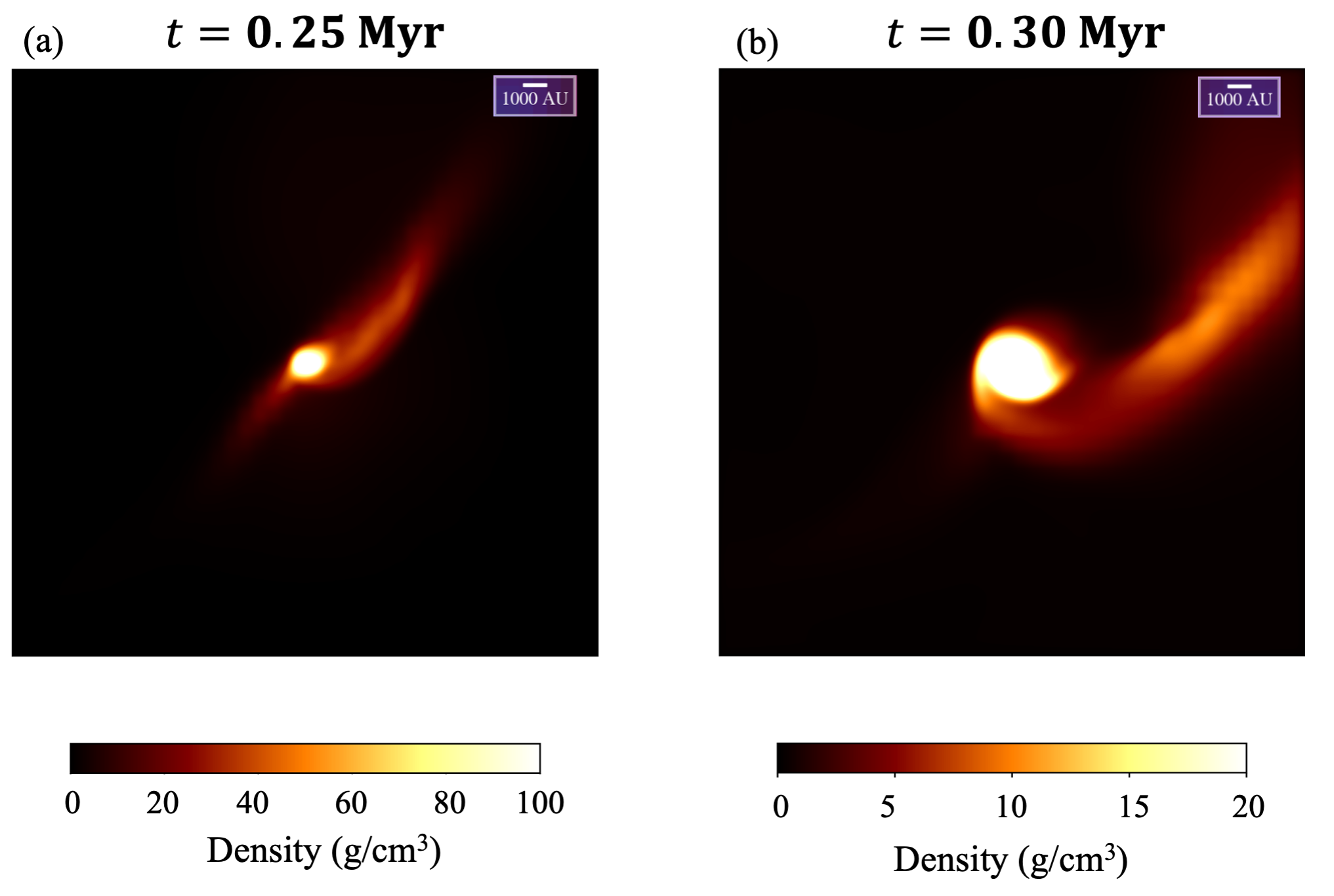} 
    \caption{Column density distributions of the main core at (a) 0.25 Myr and (b) 0.30 Myr. The images are smoothed with a Gaussian beam of 1000 au. }
    \label{fig:comparison}
\end{figure}

In Figure \ref{fig:comparison}, we show the column density distributions at $t=$ 0.25 Myr and 0.30 Myr.
In these images, we smoothed the column density distributions with a Gaussian beam to match the spatial resolution of \citet{pineda20}. The length scale of 1000 au is indicated at the upper-right corner.  The box sizes are comparable to that of Figure 1 of \citep{pineda20}.
The column density distribution presented in our model does not closely resemble the molecular line emission distribution reported by \citet{pineda20}. However, if we combine the distributions of HC$_3$N ($J=10-9$) and N$_2$H$^+$ ($J=1-0$) emissions, the resulting distribution exhibits a similar morphology to the column density distribution depicted in Figure \ref{fig:comparison}. Considering that HC$_3$N ($J=10-9$) and N$_2$H$^+$ ($J=1-0$) have similar critical densities in the range of $10^{4-5}$ cm$^{-3}$, the dense gas is expected to be distributed in a similar manner as the combined distribution. In the context of the core collision model, the gas traced by N$_2$H$^+$, which primarily distributes around the central protostar, predominantly originates from the main core, while the streamer is from the second core. Also, the discrepancy in column density between our model and the observed emission may be partly attributed to the simplicity of our numerical model, which assumes a non-turbulent and non-magnetized spherical core with a Bonner-Ebert density profile. 
The above discussion needs to be validated based on quantitative calculations of chemical evolution.

\begin{figure}[htbp]
\includegraphics[width=\linewidth]{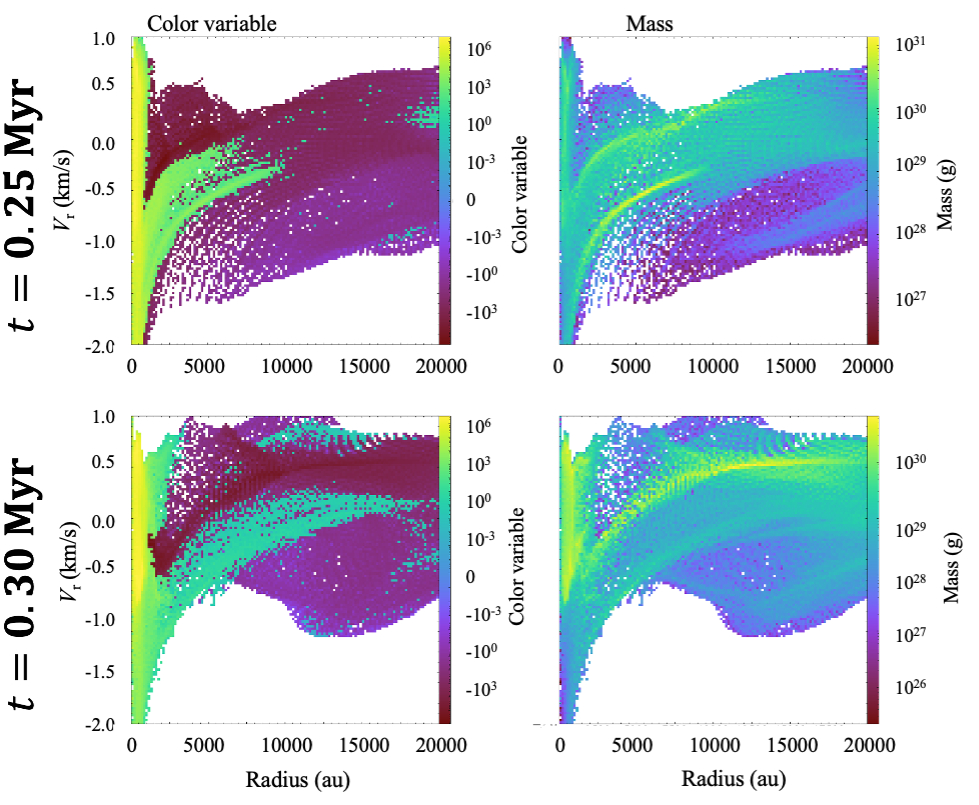}
    \caption{Phase plots of the radial velocity and color variable as a function of the distance from the sink particle. The left and right panels are the phase plots of the color variable, $C$, and mass, respectively. 
    The upper panels are the plots at $t=0.25$ Myr, while the lower panels are at $t=0.30$ Myr.}
    \label{fig:comparison2}
\end{figure}

Figures \ref{fig:comparison2} indicate the infall along the streamer. 
We note that our model shown in the present paper do not take into account the magnetic field. 
The magnetic field tends to transfer the orbital angular momentum, and thus the disk size and kinematics is likely to somewhat change in the presence of magnetic fields \citep[see also][]{kinoshita22}.

\citet{sanhueza21} observed the high-mass protostellar core associated with the spirals. They found that the two spirals have coherent velocity gradients and the velocities appears to converge to the central protostar. In addition, the magnetic fields follow the spirals.  \citet{kinoshita22}
performed the MHD simulation of core collision and they show that 
the model with weak magnetic field 
reasonably reproduces the observed density, magnetic field, and velocity structures.
Similar spiral structures also reported by \citet{xu23}.

Another compelling illustration of core collision phenomena is observed in the case of L1521F \citep{tokuda14}. 
This region exhibits multiple components with distinct velocities on a scale of 0.05 pc. Moreover, within the innermost region, high-density compact cores associated with an arc-like structure have been identified on a smaller scale of 10$^3$ au \citep{tokuda14}.
Particularly, the temperature of the arc is slightly higher than that of the core, suggesting an ongoing dynamical interaction.

It is worth noting that similar streamers can be formed by the process of ambient gas accretion, which tends to occur along the orientation of the parent filament when cores are formed through filament fragmentation. However, in the case of core collision scenarios, the formation of streamers is not strictly bound to the large-scale filamentary structure. This distinction suggests that large-scale mapping observations can serve as a valuable tool for differentiating between these two scenarios.

Furthermore, the core collision scenario predicts the generation of supersonic shocks, which could lead to the enhancement of molecular abundances, such as SO, CH$_3$OH, and SiO, along the streamers. To provide more quantitative constraints on the formation mechanisms of streamers, it will be necessary to conduct chemical evolution calculations.

In our simulations, to simplify the analysis, we assumed a zero initial angular momentum for each core. However, real cores typically possess a small amount of (spin) angular momentum prior to gravitational contraction. The origin of these angular momenta are believed to be cloud turbulence \citep[see e.g.,][]{chen18,kinoshita23}. These initial angular momenta contribute to the formation of circumstellar disks around protostars.

In the core collision scenario, orbital angular momentum plays a crucial role in the creation of circumstellar disks, which may have orientations differing from the original circumstellar disk rotation axis determined by the core's angular momentum. This discrepancy can cause complex interaction and mass loading around the central region of the protostar. For example, if the orbital and core angular momentum axes are opposite, their interaction can lead to a sudden accretion burst as the dynamical support provided by the centrifugal force is partially lost in a dynamical time. Additionally, the orientation of the outflow axis can change over time as the second core mass accretes.
These influences are likely to affect the dynamics of actual streamers.


\subsection{Stability of the circumstellar disk formed by the core collision}

In our simulations, we demonstrated that the circumstellar disk is formed by the core collision. Figure \ref{fig:tangential} shows that the rotation velocity seems to well match the Keplerian velocity calculated with the mass of the central sink particle.  Therefore, the structure near the center is rotationally-supported.
To further examine the dynamical state of the disk, we calculate the Toomre Q value.

\begin{figure}[htbp]
\includegraphics[angle=0,width=\linewidth]{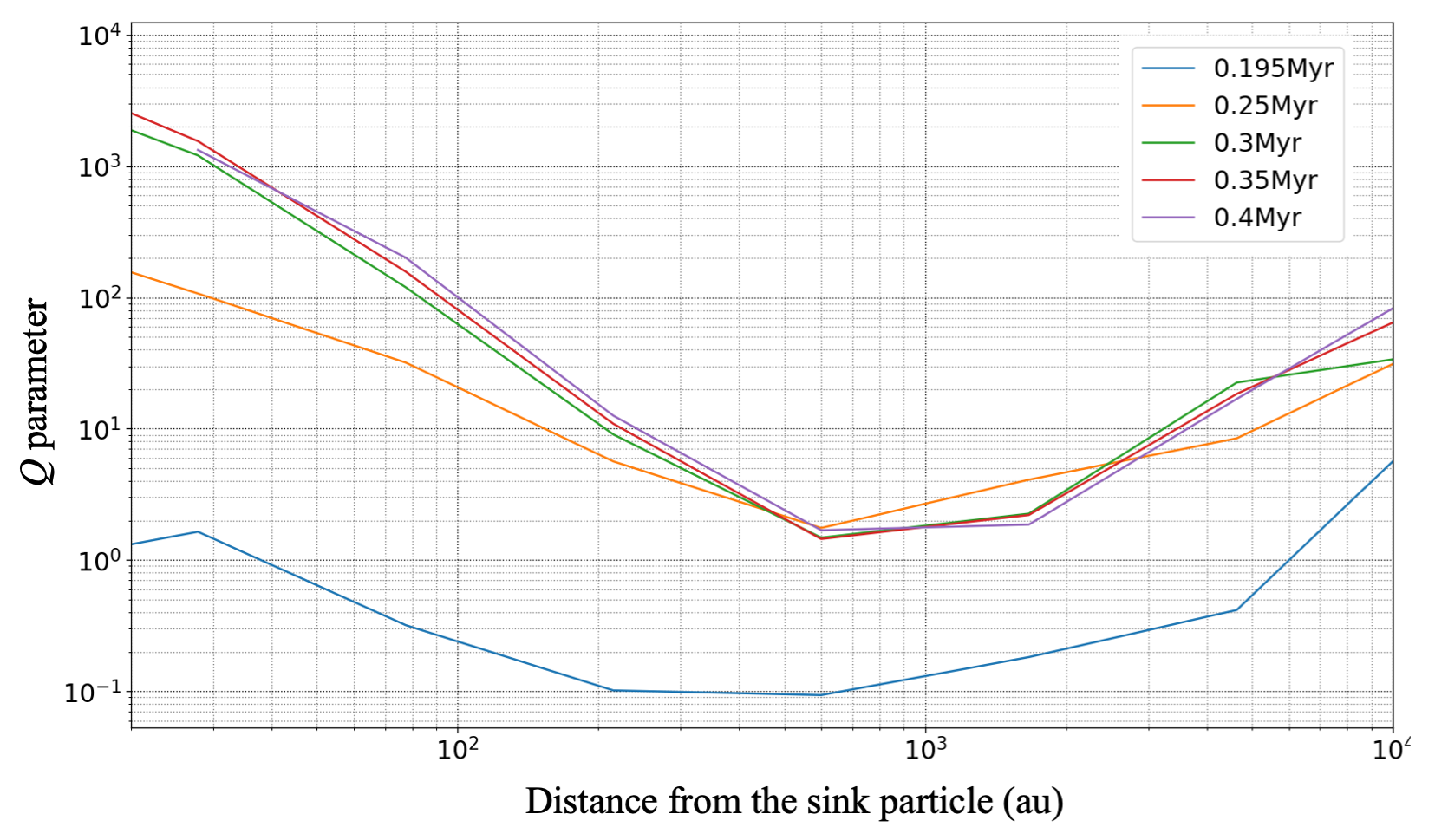}
    \caption{The distribution of the Q-value for the A5B05 model at $t=$ 0.195, 0.2, 0.30, 0.35, and 0.40 Myr.}
    \label{fig:q-value}
\end{figure}

Figure \ref{fig:q-value} shows the Q value as a function of the distance from the central sink particle, for the fiducial model A5B05.  The Toomre Q value is given as
\begin{equation}
    Q={c_s \kappa \over \pi G \Sigma_{\rm disk}}
\end{equation}
where $\kappa$ is the epicyclic frequency and in our case, 
it is equal to the angular speed.
When $Q>1$, the disk is gravitationally-stable, whereas
for $Q<1$, it becomes unstable. For $Q\sim 1$, the disk gravitational instability tends to form the spiral pattern.
The images in the previous sections show that the disks have spiral-like structures, which is reminiscent of the substructures created in the unstable disk. 
However, the computed Q-value for our disk is larger than unity over the entire disk. This indicates that the spiral structure does not originate from the disk instability. This structure is always connected to the large-scale streamers and is presumably comes from the infalling gas stream having angular momentum, not from the gravitational instability of the rotating disk. 
The Q-value becomes somewhat lower at around $r\sim 500$ au, as time goes by. 
Once the sufficient amount of gas accretes onto the central part, the disk will eventually become self-gravitating.

In our numerical model, the core's initial rotation was set to zero. However, real cores have spin angular momentum before their contraction, which leads to the formation of a rotating circumstellar disk.  Our simulation show that the core collision setup has the orbital angular momentum between the two cores which can be used to form the other circumstellar disk whose rotation axis can differ from the rotating disk formed by the direct gravitational collapse.
Their interaction may significantly influence the evolution of the rotating disks, and may change the stability condition discussed above.

\subsection{Variability of the mass accretion}

\begin{figure}[htbp]
\includegraphics[width=\linewidth]{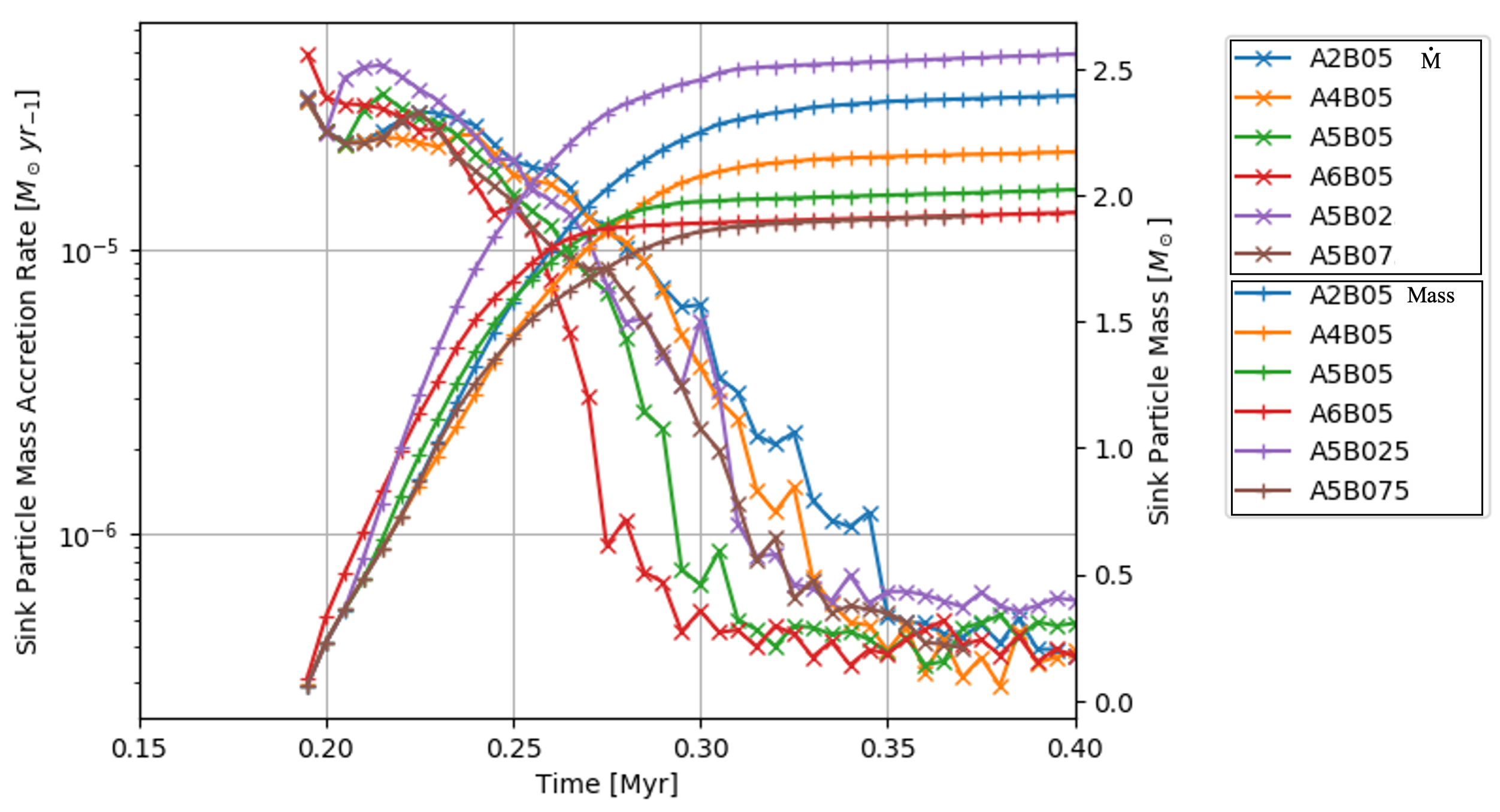}
    \caption{Time evolution of mass accretion rate and the central sink mass.}
    \label{fig:accretion}
\end{figure}

The core collision tends to form asymmetric structure around the circumstellar regions. Such asymmetric structure is likely to affect the mass accretion toward the central protostar.  In our simulations, such effect can be seen in the time evolution of the mass accretion rate. 
In Figure \ref{fig:accretion} we show the time evolution of the mass accretion toward the central protostar or sink particle.  
The mass accretion takes its maximum immediately after the sink particle formation.
Then it declines with time.  Such a behavior is qualitatively similar to that of the isothermal spherical collapse. 
The actual maximum rate depends on the spatial resolution of the simulations. In the fiducial model A5B05, about 80 \% of the main core mass goes into the sink particle by $t\sim 0.27$ Myr.
After 0.30 Myr, the mass accretion rate appears to approaches to a constant value of $3\times 10^{-7}$ $M_\odot $ yr$^{-1}$.  
This behavior is different from the spherical collapse. For the single spherical collapse, the accretion rate takes its maximum right after the sink formation and declines with time \citep{larson69, foster93, ogino99}. 
The mass accretion rate also varies with about 50 \% level. This fluctuation may be generated from the coiled motions observed in the disk.

\subsection{Binary formation}

We showed in Section \ref{subsec:speed} that for the model with a relative speed of $V_{\rm col}=7 c_s$, a binary star was formed. 
The collision speed is about 1.4 km s$^{-1}$ under the assumption of $T=10$ K. This is comparable to the velocity dispersion observed in typical star-forming clouds.
The separation is about 5000 au at the formation stage.
However, the second star do not have enough tangential velocity, it fall toward the primary star, and the separation becomes small at $<10^2$ au.  

The faster collision speed leads to larger compression, so that the density in the compressed layer becomes higher. Such a dense part in the compressed layer collapses to form a less-massive star. 
In this triggered binary formation, the binary mass ratio is likely to be deviated from unity.
These stars also have the streamer-like structure.

The formation of binary stars is also expected to depend on the initial model parameters such as the core masses and densities. But, we expect that binary or multiple stars can be formed easily for the faster collision speed.

As discussed in Section \ref{sec:frequency}, the collision frequency is relatively high within clustered environments. Furthermore, based on the line-width-size relations, the observed 3D velocity dispersion of the molecular gas often reaches magnitudes of a few km s$^{-1}$. Considering these factors, the occurrence of binary formation triggered by core collisions emerges as a prominent process for the formation of binary systems within clustered environments.

An interesting aspect of binary systems formed through core collisions is the potential for different rotation axes. This arises from the fact that the direction of the orbital angular momentum can differ from the spin angular momentum of the gravitationally-contracting core. Therefore, it is possible for a binary system to exhibit misaligned rotation axes or counter-rotation.

\subsection{A brief Comparison with \citet{kinoshita22}'s MHD simulations}

In this paper, we have not taken into account magnetic fields, which are known to have a significant impact on the dynamical evolution of clouds and cores. However, it is important to briefly discuss the potential effects of magnetic fields on the core collision processes. \citet{kinoshita22} conducted MHD simulations involving two initially stable, identical cores. Their results demonstrate that core collisions can induce gravitational contraction of the cores by shock compression, similar in qualitative nature to the non-magnetized scenario discussed in our study.

Considering the profound influence of magnetic fields on gas dynamics, we anticipate several possible important influences if we include magnetic field in our simulations.
Firstly, the presence of magnetic fields leads to a reduced dynamic compression during collisions, resulting in a less pronounced density enhancement compared to the non-magnetized case. Formation of the circumstellar disks is delayed. Moreover, magnetic fields tend to align with the streamers, as the magnetic field components tangential to the shock plane are preferentially amplified.
Secondly, the efficient magnetic braking effect amplifies the significance of inflow along the streamers, potentially leading to more rapid mass accretion onto the central sink particle.
These characteristics were observed in the magnetized core collision conducted by \citet{kinoshita22} (see e.g., their figures 17 for the unequal-mass core model).
Therefore, it is crucial to account for these magnetic field-related effects in order to obtain a comprehensive understanding of the core collision processes and streamer formation. In future, we will discuss the effect of magnetic field on the core collision in more details.

\clearpage
\section{Summary}
\label{sec:summary}

In the present paper, we discussed the importance of the collision between two dense cores in star formation process.  The main results are summarized as follows.

\begin{itemize}
\item[1.] We estimated the collision frequencies in several star-forming regions using the published core catalogs and fragmentation scenario. We found that the collision frequency is larger than unity in most  star-forming regions.  
Even for fragmentation scenario, the core should collide with adjacent cores in its lifetime.
Therefore, we conclude that the core collision is likely to be an important process in star formation.

\item[2.] We carried out the hydrodynamical simulations of the collision between unequal-mass cores. One core is gravitationally unstable, and the other is stable.
We showed that the core collision creates the rotating disk associated with a prominent curved arm.
Such a structure morphologically resembles the observed streamers and spirals recently found around the protostellar systems \citep[e.g.,][]{pineda20,chen21,sanhueza21,xu23,hsieh23}.

\item[3.]
When the collision speed is large, another protostar is formed in the shock-compressed layer, and thus a binary system is formed.
Although the separation is about 5000 au at the formation epoch, it becomes smaller at about $10^2$ au.

\item[4.] The shock-compressed gas tends to fall onto the central protostar along the arm or streamer.
Such inflow along the arm or streamer creates a coiled density structure around the protostar.
When the collision speed is large, the outer part of the streamer was torn away.

\item[5.] The core collision causes the time-variable accretion onto the central protostar. 

\end{itemize}


This work was financially supported by JSPS KAKENHI Grant Number JP23H01218.
\bibliography{nakamura}{}

\end{document}